\documentclass{elsart}
\usepackage{epsfig}
\begin{document}

\begin{frontmatter}
\title{Nanoscopic stripe-like inhomogeneities
                                     and optical conductivity of doped cuprates}
\author[usu]{A.S. Moskvin, E.V. Zenkov, Yu. D. Panov}
\address[usu]{Ural State University, Ekaterinburg, 620083,  Russia}

\begin{abstract}
We propose a new approach to theoretical description of doped cuprate like \\
$La_{2-x}Sr_{x}CuO_4$ and $YBa_2Cu_3O_{6+x}$, assuming phase separation and
treating it as inhomogeneous composite material, containing the dielectric and
metallic stripe-like nanoparticles. The formalism of effective medium theory is
then applied for calculation of dielectric permittivity, optical and EELS
spectra of $La_{2-x}Sr_{x}CuO_4$ with $x$ varying in a wide range. Reasonable
semi-quantitative agreement with experiment has been obtained even for the
simplest version of the theory. The model was found able to reproduce all
essential features of optical conductivity $\sigma(\omega)$  and transmittance
both for thin films (M. Suzuki, Phys. Rev. {\bf B 39}, 2321 (1989)) and bulk
single-crystalline samples (S. Uchida {\it et al.} Phys. Rev. {\bf B 43}, 7942
(1991)). Substantial difference in spectral and doping dependence of optical
absorption for the thin-film  and bulk samples is easily explained if only to
assume different shape of metallic and dielectric regions in both materials.
New peaks in $\sigma(\omega)$ and absorption spectra, that emerge in the
midinfrared range near $0.5$ and $1.5$ eV upon doping are attributed to
geometrical (Mie's) resonances. Overall, we point out that  all main
peculiarities of the doping effect on optical and EELS spectra for cuprates
including the spectral changes accompanying the insulator-to-metal transition
can be explained rather prosaically  by recognizing that the doping results in
emergence of nanoscopic metallic stripe-like droplets.
\end{abstract}
\end{frontmatter}

\section{Introduction}
High-$T_c$ superconductivity and other unconventional properties of doped
quasi-2D cuprates remain a challenging and hot debated  problem. As usual, the
active $CuO_2$ planes in cuprates are considered as homogeneous systems,
starting from one-, three- or multiband Hubbard models. Often only in-plane
$Cu(3d)-O(2p\sigma)$ basis set is taken into account and the well defined
Zhang-Rice singlet ground state for doped holes is presumed. In frames of such
theories the dependence of various properties of copper oxides on doping are
related to the rearrangements of Hubbard bands.

However there are clear experimental evidences in favour of intrinsic and
generic  inhomogeneous nature of the systems under consideration with well
developed static and dynamic spatial inhomogeneities \cite{Phillips}.
Nonisovalent chemical substitution in insulating cuprates like
$La_{2-x}Sr_{x}CuO_4$ and $YBa_2Cu_3O_{6+x}$ results in an increase of the
energy of the parent phase and creates proper conditions for its competing with
other, possibly metallic phases capable to provide an effective screening of
the charge inhomogeneity potential.

At the beginning (nucleation regime) a new phase appears in form of a somewhat
like a metallic droplet in insulating matrix. The parent phase and droplets of
new phase may coexist in the phase separation regime. At different stages of
the "chemical" phase separation regime we deal with the isolated droplets, with
the percolation effect and finally with a complete removal of the parent phase
\cite{Moskvin}. This viewpoint on the phase separation phenomena in cuprates is
generally compatible with the pioneer ideas by Emery and Kivelson \cite{Emery}
and some other model approaches. So, on the basis of the neutron scattering
data Egami \cite{Egami} conjectured an appearance of a nano-scale heterogeneous
structure which is composed of a plenty mobile-carrier existing region of
metallic conductivity and semi-localized scarce carrier region with
antiferromagnetic spin ordering.

At present, there are numerous experimental indications on the phase separation
and nanoscopic stripe-like textures in doped cuprates. Hence when considering
the phenomena with a relevant characteristic length, e.g. optical absorption in
infrared ($\lambda \simeq 0.5 - 2\, \mu m$), doped cuprates can be roughly
regarded as a binary granular medium, composed of insulating and conducting
regions and described in frames of the formalism of an {\it effective medium}
(EM) theory.

\section{Physical properties of composites: effective medium approach}

In what concerns the optical properties the classical problem of EM is known
since long and is related to the calculation of some property of a composite
system (typically, the dielectric permittivity $\varepsilon_{eff}$), those of
pure components and their volume fractions been given. Few versions of the
theory have been proposed \cite{eff}. Many of them were shown to follow from
one integral equation \cite{Stroud} and have inherent shortcoming, assuming the
concentration of one of components to be small. Among them the Maxwell-Garnett
theory is one of the most popular ones for its simplicity. More sophisticated
version of the theory was given by Ping Sheng \cite{PShg}; it has been proven
valid in entire range of concentration of components.

Metal clusters and small metallic particles embedded in a dielectric matrix
have been extensively studied in the past years. One of the most interesting
features of these systems is an appearance of the morphological (geometrical),
or Mie resonances  \cite{Mie} related  to the excitation of surface plasmon
modes in nanoparticle when it scatters the light wave of appropriate frequency.
The dispersion law of surface eigenmodes in particles explicitly depends on its
geometric shape. Rigorous theory of this effect was first derived by Mie for
the case of a spherical particle \cite{Mie}. In principle, infinite number of
geometric resonances should be observed, that correspond to dipolar,
quadrupolar etc. surface eigenmodes. However, only dipole contribution is
practically accounted. This is valid provided that the particle is placed in
homogeneous field, i.e. that the size of the particle is small compared with
the wavelength. The frequency of geometric resonance is then easily obtained as
the one at which the polarizability of small particle diverges. For the case of
spherical particle embedded in the matrix with dielectric permittivity
$\varepsilon_d$ this leads to the equation:
\begin{equation}
 \varepsilon(\omega)_{part.}\,+\,2\,\varepsilon_d\,=\,0,
\end{equation}
whence the resonance frequency is
\begin{equation}
  \omega_r\,=\,\frac{\omega_p}{\sqrt{1\,+\,2\,\varepsilon_d}},
\end{equation}
if the Drude's expression is adopted for the particle permittivity and
$\varepsilon_d$ is assumed to be constant. In the case of arbitrary ellipsoid
there are three different principal values of polarizability and the last
formula generalizes to
\begin{equation}\label{ell}
\omega_r^i\,=\,\omega_p\,\sqrt{\frac{L_i}{\varepsilon_d\,-\,L_i\,(\varepsilon_d\,-\,1)}},
\ i\,=\,1,\,2,\,3,
\end{equation}
where $L_i$ are three (shape-dependent) depolarizaton factors. In both
expressions $\omega_p$ is the plasma frequency.

Below we outline in brief some underlying ideas of the EM theory having in mind
its further application to doped cuprates. The medium is assumed to be composed
of ellipsoidal particles of two kinds. Recalling the lamellar structure of
copper oxides we assumed the particles to be randomly oriented in {\bf ab}
plane, one of their principal axes being aligned in {\bf c} direction. When one
addresses the microstructure of such composite two cases may occur, when either
of components form the grain and other - the coating. The essential of the
theory consists in treating the structure of material as a weighted average of
these two conformations in proportion, that depends on volume fraction of
components. This has an advantage in correctly displaying the percolation
threshold. Final expression which determines an effective dielectric function
has a form \cite{PShg}:
\begin{equation}
   f\,D(\varepsilon_{eff},\,\varepsilon_1,\,\varepsilon_2,\,p)\;+
\;(1\,-\,f)\,D(\varepsilon_{eff},\,\varepsilon_2,\,\varepsilon_1,\,1\,-\,p)\;=0\,.
\end{equation}
Here, the quantity $D(\varepsilon_{eff},\,\varepsilon_1,\,\varepsilon_2,\,p)$
is the orientationally averaged dipole moment of a kind-I particle coated by
kind-II particles, embedded in effective medium, $f$ and $(1-f)$ are relative
probabilities of occurrence  for two kinds of microstructure of the composite.
Rough estimate yields $f$ in the form: $f\,=\,v_1\,/\,(v1\,+\,v2)$,
$v_1\,=\,(1\,-\,\sqrt[3]{p})^3$, $v_2\,=\,(1\,-\,\sqrt[3]{1\,-\,p})^3$, where
$p$ is the volume fraction of component I.  Analytical derivation of
corresponding expression needs the grain and coating to be confocal and leads
to the result \cite{Bilboul}:
\begin{eqnarray}
D(\varepsilon_{eff},\,\varepsilon_1,\,\varepsilon_2,\,p)& = & \nonumber \\ {1
\over 2} \sum_{i\,=\,1,\,2}\,{p\,(\varepsilon_1 - \varepsilon_2)\,
\varepsilon_2\,+\,\left(\left(L_i^{in} - p\, L_i^{out}\right)\, (\varepsilon_1
- \varepsilon_2)\, +\, \ \varepsilon_2\right)\, (\varepsilon_2 -
\varepsilon_{eff}) \over p\, L_i^{in}\, (\varepsilon_1 - \varepsilon_2)\,
\varepsilon_2\, +\,\left(\left(L_i^{in} - p\, L_i^{out}\right)\, (\varepsilon_1
- \varepsilon_2)\,+\,\varepsilon_2\right)\, \left(\varepsilon_{eff}\, -\,
        L_i^{out}\, (\varepsilon_{eff} - \varepsilon_2)\right)} ,
\label{coat}
\end{eqnarray}
where $L_i^{in,\,out}$ are two in-plane depolarization factors of inner and
outer ellipsoids, respectively. These are defined as follows:
\begin{equation}
  L_i\,=\,\frac{a_1\,a_2\,a_3}{2}\,\int_0^{\infty}\frac{d t}{(t\,+\,a_i^2)\sqrt{\sum_{k=1}^3
  (t\,+\,a_k^2)}},
\label{depol}
\end{equation}
and  can be easily re-expressed in terms of ratios between the semi-axes of
ellipsoid.

Despite the shape  anisotropy of nanoscopic inclusions the dielectric fuctions
for both media I and II together with the effective medium are assumed to be
isotropic for a sake of simplicity. This assumption of optically isotropic
media implies only a semi-quantitative description of such an anisotropic
system as quasi-2D cuprate.

\section{Optical spectroscopy of doped cuprates: experiment and theoretical models}

Optical properties of HTSC cuprates have been the subject of numerous
experimental and theoretical  investigations. One of the first systematic
studies of the doping dependence of the optical reflectance and transmittance
for {\it epitaxial films} of $La_{2-x}Sr_xCuO_4$ deposited on $SrTiO_3$
substrate were reported in \cite{Suzuki}. The most dramatic change with
increasing $x$ was observed near $1.5$ eV.  The author speculated that the
emergence and near proportional strengthening of the additional absorption
centered at $1.5$ eV (see Fig. \ref{suzuki})  might be related to a dramatic
change in the valence band structure. However, the careful measurements of
optical reflectivity spectra for {\it bulk single crystals} of
$La_{2-x}Sr_xCuO_4$ with $x$ varying in wide range, which covers the whole
phase diagram of the material, performed by Uchida {\it et al}. \cite{Uchida},
reveal only a relatively weak  feature near $1.5$ eV which becomes apparent
only for $x>0.10$. The authors suppose that the $1.5$ eV feature is extrinsic
in origin, and may be related, for instance, to some undetectable amount of
oxygen vacancies.

Main effect of doping in $La_{2-x}Sr_xCuO_4$ and other cuprates is associated
with the gradual shift of the spectral weight from the high-energy part of the
spectrum to the low-energy part with formation of the prominent midinfrared
(MIR) region absorption band and  pronounced Drude-like peak. This spectral
shift is particularly emphasized by the presence of the {\it isosbestic} point
at which $\sigma (\omega)$ is invariant against $x$ \cite{Uchida}.

The authors \cite{Suzuki,Uchida} state that simple band theory cannot describe
the low-energy part of the spectrum within $CuO_2$ planes, whereas it nicely
mimics the higher-energy excitations. For $x>0.05$ the former is clearly
separated into a Drude-type peak at $\omega =0$ and a broad band centered in
the mid-IR region.

One of the puzzling features of the reflectivity spectra in $La_{2-x}Sr_xCuO_4$
and other cuprates is that the position of the  reflectivity  edge is almost
unchanged against dopant concentration and eventually becomes a plasma edge of
the Drude carriers in the overdoped region. In frames of a conventional
approach it means that the edge frequency, or a screened plasma frequency, has
nearly constant magnitude that rules out the simple relation $n\sim x$, which
means that the edge is associated straightforwardly with the doped holes.
Instead, it is plausible to suppose that $n\sim (1-x)$ \cite{Uchida1}.
Comparative analysis of the reflectivity spectra of various cuprates
\cite{Uchida1} shows that the plasma frequency of the materials with a nearly
half-filled band is almost independent of $x$ for small doping but dependent on
the average spacing of the $CuO_2$ planes. This conclusion clearly contradicts
to the results based on the $t-J$ model \cite{Horsch} which show a rather
strong dependence of plasma frequency $\omega_p$ on hole concentration,
especially at small doping.

Extremely lightly oxygen doped crystals $La_2CuO_{4+y}$ grown by two different
methods (top-seeded and float-zone) \cite{Kleinberg} exhibit different
low-temperature  optical absorption spectra from $0.1$ to $1.5$ eV photon
energy, albeit at room temperature the spectra are nearly identical both each
other and those for slightly "chemically" doped $La_{2-x}Sr_xCuO_4$
\cite{Uchida} and "photo-doped" $La_2CuO_{4}$ \cite{Perkins1}. In other words,
in all cases, irrespective of the "external" cause, we deal with the same
intrinsic mechanism of optical absorption.

Electron-hole optical excitations for the doped cuprate have been calculated by
Wagner {\it et al}. \cite{Wagner} via exact diagonalization on finite clusters
for a 2D multiorbital Hubbard model. Doping with one and two holes for 12-site
lattices introduces a shift of absorption weight from the high-energy side to
the low-energy one, in general agreement with experiment. Three additional
features can clearly be identified on the low-energy  side: first, the Drude
peak with weight $\propto x$, second, a small feature caused by internal
transitions in the Zhang-Rice singlet manifold which may be related to the
experimentally observed midinfrared absorption. Finally, a strong absorption
structure caused by the transitions from the Zhang-Rice singlet states into the
upper Hubbard band was detected and assigned to experimental spectral weight
found in the conductivity measurements between the Drude and the charge
transfer peaks.

The optical conductivity has been calculated using the Hubbard and $t-J$ models
by exact diagonalization on small clusters \cite{Dagotto}. These studies
predict that at low hole density, in addition to a Drude peak with a spectral
weight proportional to doping there is a MIR band centered at photon energy
$\sim 2J$, which should be rather broad and presumably caused by hole coupling
to spin excitations. The high energy ($\geq 4J$) of the MIR bands as observed
experimentally is in clear contradiction with such a theory.

Later on Lorenzana and Yu \cite{Lorenz} calculated the optical conductivity of
$La_{2-x}Sr_xCuO_4$ in the inhomogeneous Hartree-Fock plus random-phase
approximation using the $p-d$ model with parameters taken from first-principle
calculations. The same as for Wagner {\it et al.} \cite{Wagner} CT band peaks
now  at $1.7\, t_{pd}\approx 2.7$ eV. The precursor of the MIR band is
dentified with transitions from the localized states on $Cu$ to the extended
states above the Fermi level. Despite the authors \cite{Lorenz} state close
agreement with experiments regarding the peak positions and relative
intensities for zero and small doping, their model  calculations fail to
properly describe the overall evolution of  spectra with dielectric-to-metal
transition.

\section{The effective-medium theory of optical conductivity in doped cuprates}

We have applied the formalism of effective medium theory \cite{PShg} in order
to reproduce some tendencies in evolution of optical spectra of high-$T_c$
cuprates upon doping. One should note that the optical conductivity and its
doping dependence are considered to be one of the most important and
informative characteristics of the doped cuprates.
\begin{figure}
\centering{\epsfig{file=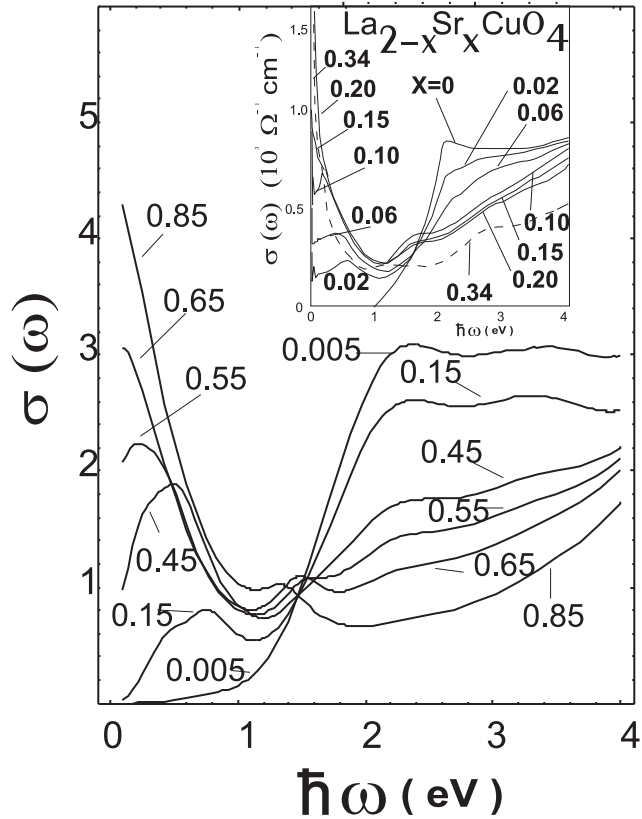,width=20pc}} \caption{Calculated optical
conductivity for the model bulk single-crystalline $La_{2-x}Sr_xCuO_4$ system
at different volume fractions of metallic phase. Inset: measured optical
conductivity \cite{Uchida} for $La_{2-x}Sr_xCuO_4$ at different $x$.}
\label{uchida}
\end{figure}
In particular, we have attempted to describe the experimental data on doping
dependence of optical conductivity in $La_{2-x}Sr_xCuO_4$ \cite{Suzuki,Uchida}
in framework of the EM model. It should be emphasized that in our case the EM
theory is nothing but a phenomenological approach to the description of the
complex spatially non-uniform  rearrangements in the electronic structure of
doped cuprate. The concept of binary composite merely reflects the fact that
upon doping some fraction of electrons remains localized in the regions of
parent insulating phase while other electrons becomes nearly metallic inside
regions embedded into parent insulating matrix.

It is reasonable to conjecture, that different electronic states of
La$_2$CuO$_4$ do not contribute equally to this process. In particular, we
simply assumed that CT excitation band(s) peaked near $2.5$ eV is involved in
such a "metallization" while the states at higher energy remain rigid. These
ideas can easily account for the behaviour of optical conductivity of
La$_{2-x}$Sr$_x$CuO$_4$: upon doping the spectral weight rapidly transfers from
the CT band(s) to a narrow band of extended states, while the high-energy tails
converge beyond $\sim\,4.5$ eV for different $x$.

The imaginary part of $\varepsilon$ for the parent undoped cuprate $La_2CuO_4$
was taken from \cite{Uchida} and \cite{Falck}. The spectrum of
Im\,$\varepsilon$ of $La_2CuO_4$ we used in our calculations was modeled by a
superposition of two CT bands fitted by a sum of two Gaussians centered at
$2.0$ and $3.0$ eV, and the "rigid" high-energy contribution fitted by a single
Lorentzian centered at $6.0$ eV. Only the first CT term is involved in EMT
computations, its spectral weight being partly substituted by that of new
metallic-like phase. As for the real part of dielectric function, near
dispersionless value close to $\epsilon =\,4$ has been chosen.
\begin{figure}
\centering{\epsfig{file=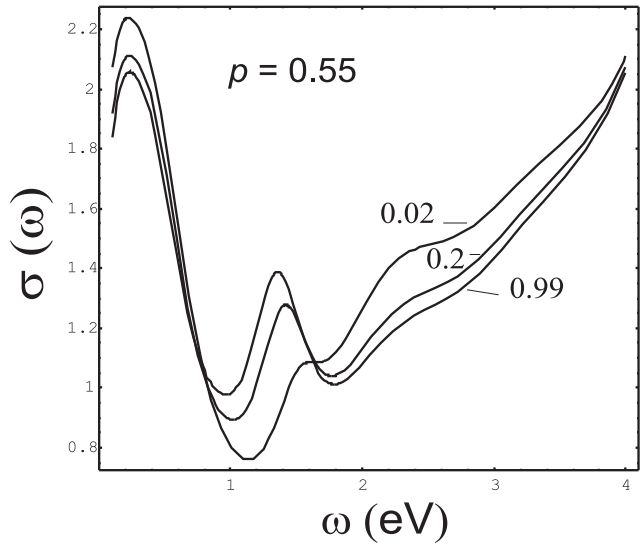,width=20pc}} \caption{Optical conductivity
for the model $La_{2-x}Sr_xCuO_4$ system at different shape of inclusions
(lesser number corresponds to more plate particles embedded in {\bf
ab}-plane).} \label{shape}
\end{figure}
To model the dielectric function of the metallic phase we used the simplest
Drude form
\begin{equation}\label{Drude}
  \varepsilon(\omega)\;=\;\varepsilon_{\infty}\,\left(1\ -\
  \frac{\omega_p^2}{\omega\,(\omega\,+\,i\,\gamma)} \right),
\end{equation}
where $\omega_p$ and $\gamma$ are the plasma frequency and  the electron
scattering rate, respectively. Together with $\varepsilon_{\infty}$ these are
regarded as adjustable parameters which magnitude does not depend on doping,
frequency, and temperature. The results presented below were obtained with the
values: $\omega_p\,=\,1.65$ eV, $\gamma\,=\,0.6$ eV,
$\varepsilon_{\infty}\,=\,1.0$, which once tuned, were retained unchanged
against the doping level.

Another set of free parameters stand for the microtexture of the sample under
investigation. Here we take into account numerous manifestations of stripe-like
textures in doped cuprates. We denote the in-plane semi-axes ratio for
particles of both kinds as $\alpha$ and that of out-of-plane and major in-plane
semi-axes as $\beta$. These geometrical parameters enter  the theory through
the depolarization factors  (\ref{depol}) and were found to affect strongly the
dielectric function of doped cuprate.
\begin{figure}
\centering{\epsfig{file=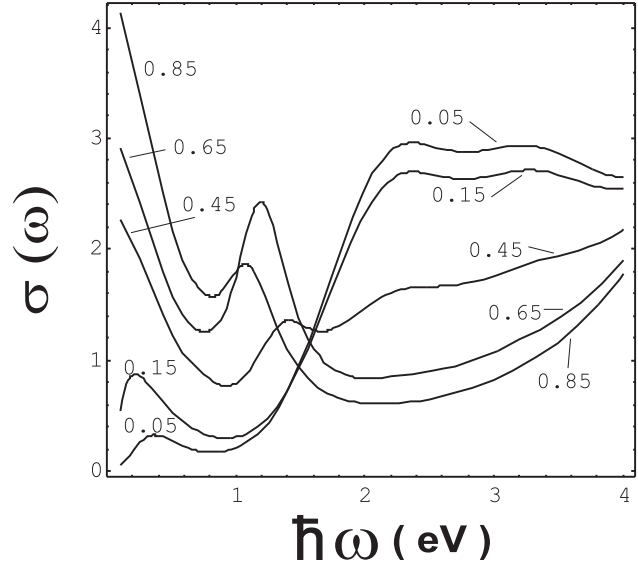,width=20pc}} \caption{Calculated optical
conductivity for the thin film $La_{2-x}Sr_xCuO_4$ sample at different volume
fractions of metallic phase:  effect of the modified
 particle shape (see text).}
\label{uchid2}
\end{figure}

Calculated spectra of optical conductivity $\sigma(\omega)$ for different level
of doping $x$ are shown in Fig. \ref{uchida} together with experimental curves
from \cite{Uchida}. The numbers near the curves stand for the fractions of
metal phase $p$, which are functions of $x$. The geometrical parameters
$\alpha_1\,=\,0.3,\;\beta_1\,=\,0.09$ and $\alpha_2\,=\,1.0,\;\beta_2\,=\,0.05$
were adopted for metallic and insulating grains, respectively. This implies the
effective metallic nanoparticle to be stripe-shaped (elongated in {\bf ab} -
plane and nearly plate in the {\bf c} - direction) and dielectric ones to be
oblate ellipsoid-like round pellets (equal in-plane semi-axes and much smaller
out-of-plane semi-axis). Spectral weight was found to transfer dramatically
from CT band to a low-energy band at $0.7$ eV in the range $p\,\sim\,0.3$ that
corresponds to $x\,\sim\,0.06$. Further increase of metal fraction results in
full removal of $2.2$ eV feature of parent compound and emergence of a new weak
peak at $\sim\,1.5$ eV for large enough $p$, really seen in the heavily doped
compositions ($x\,\sim\,0.15$). Simultaneously, the $0.7$ eV band shifts to
lower energies and gradually transforms into the Drude-like peak. Many doped
cuprates demonstrate isosbestic behaviour of optical conductivity, particularly
well pronounced e.g. in $T^{\prime}$ phase of $Nd_{2-x}Ce_{x}CuO_4$:
$\sigma(\omega,\,x)$ cross the same point for various $x$. It is seen that our
model theory remarkably well reproduces this feature.

Overall agreement of simple EM theory with experimental data \cite{Uchida} is
rather impressive despite such oversimplified assumptions as  invariable value
of shape and size of grains, plasma frequency, relaxation rate and other
quantities regardless the doping level. At the same time one observe clear
shortcomings in what concerns the detailed quantitative description of the
low-frequency ($\hbar \omega \leq \gamma$) spectral range and the percolation
phenomenon. The EM theory yields the percolation threshold volume fraction $p_c
\sim 0.6$ that seems too high. The more detailed description of the
low-frequency spectral range in frames of the EM theory first of all implies
the account of the size and shape distribution of metallic/dielectric grains
along with the distribution of quantities like $\omega _{p}$ and $ \gamma$.
\begin{figure}
\centering{\epsfig{file=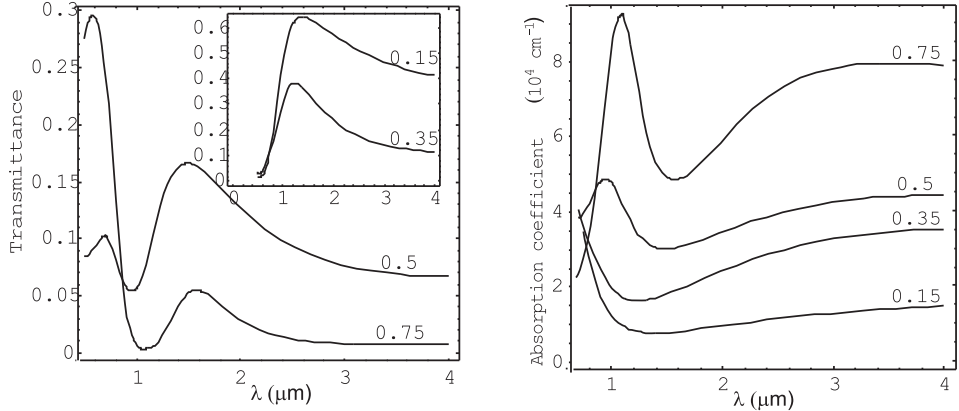,width=30pc}} \caption{Calculated absorption
coefficient (right panel) and transmittance (left panel) for the thin film
$La_{2-x}Sr_xCuO_4$ sample vs. wavelength at different volume fractions of
metallic phase.} \label{suzuki}
\end{figure}
We have identified the low-energy peak and $1.5$ eV feature with geometric Mie
resonances  as they coincide with divergencies in some terms of polarizability
of composite (nullification of denominator in (\ref{coat})). This is a quite
novel type of resonant absorption, inherent for composite materials, which is
not originated straightforwardly from either electronic transitions. The strong
influence of the grain shape on the $1.5$ eV feature is fairly well illustrated
in Fig. \ref{shape} which demonstrates the calculated optical conductivity
$\sigma(\omega)$ given the metallic volume fraction $p=0.55$ at different
values of parameter $\beta$ which defines the ratio of the out-of-plane and
in-plane semi-axes of dielectric grain.
\begin{figure}
\centering{\epsfig{file=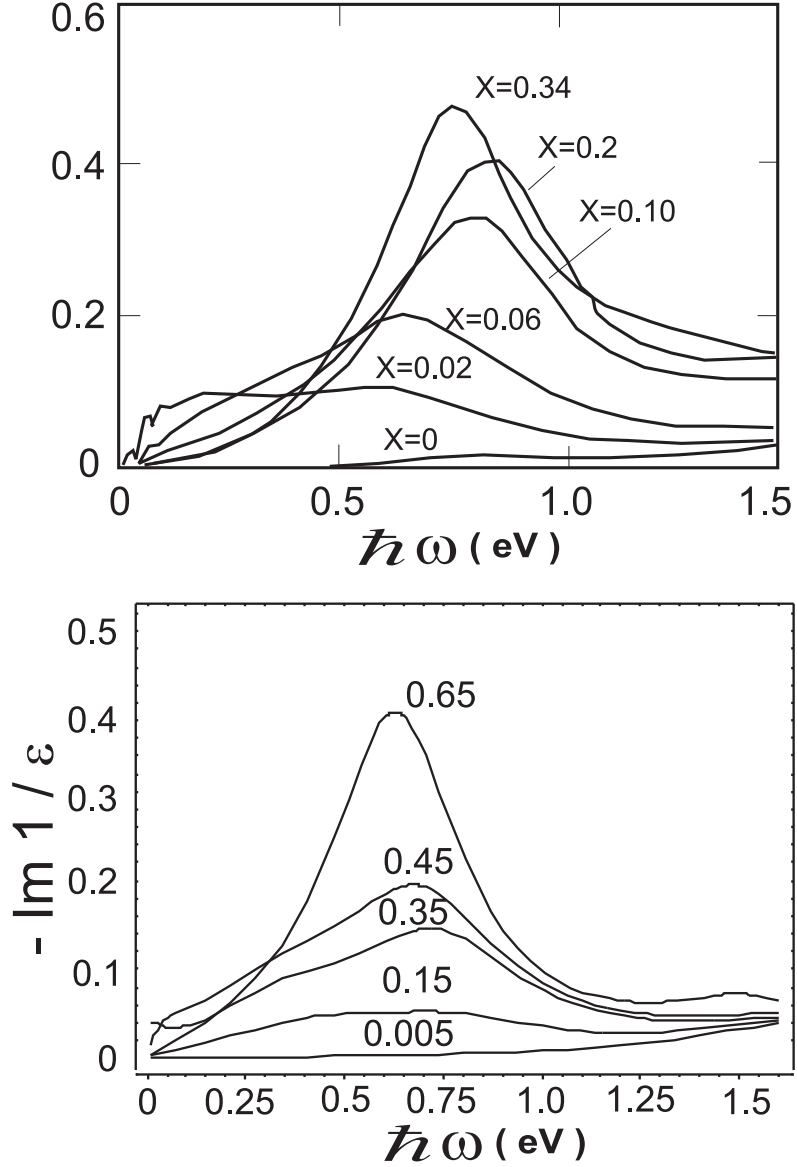,width=25pc}} \caption{EELS spectra of
$La_{2-x}Sr_xCuO_4$ at $q=0$: experiment \cite{Uchida} (upper panel) and the
model EM theory (lower panel).} \label{eels}
\end{figure}

The prominent  $\sim\,1.5$ eV peak is substantially more pronounced in optical
absorption spectra reported by Suzuki \cite{Suzuki} for thin-film samples. We
show that good agreement with both sets of the experimentals
\cite{Suzuki,Uchida} could be obtained merely adjusting the "geometric", or
shape parameters of metallic and insulating droplets without modifying any
underlying physics of the model. Calculated spectra of optical conductivity
$\sigma(\omega)$, absorption $\alpha(\lambda)$ and transmittance $T(\lambda)$
for the $0.6\,\mu$m thin film samples of $La_{2-x}Sr_xCuO_4$  are presented in
Fig. \ref{uchid2} and Fig. \ref{suzuki}. Actually, all physical parameters were
retained unvaried except of Drude's relaxation rate: $\gamma\,=\,0.4$ eV was
set instead of previous value 0.6 eV, but new values for shape parameters
$\alpha_1\,=\,0.9,\;\beta_1\,=\,0.01$ and $\alpha_2\,=\,0.9,\;\beta_2\,=\,1.5$
are taken. This implies the oblate ellipsoid-like metallic grains and
predominantly  3D character of dielectric regions. The most essential
consequence of these changes is associated with pronounced development of $1.5$
eV feature in optical conductivity. Accordingly, the $\alpha(\lambda)$ and
$T(\lambda)$ spectra fit fairly well the experimental curves \cite{Suzuki}.

Real part of optical conductivity addressed above relates to imaginary part of
dielectric permittivity. As an important additional check of validity of our
simple model we address  the calculation of the electron energy  dependent loss
function  Im$(-1/\epsilon (\omega ))$ which is determined both by real and
imaginary parts of $\epsilon (\omega )$. The Figure \ref{eels} presents a
calculated spectral dependence of the electron energy loss function for a model
$La_{2-x}Sr_xCuO_4$ system given the same model parameters used above for
description of the optical conductivity for the bulk samples (see Fig.
\ref{uchida}). Comparison with experimental data \cite{Uchida} shows that our
simple model allows correctly describe all features of the EELS spectra, in
particular, a non-monotonous doping  dependence of the peak in the loss
function. It should be noted that in contrast with conventional approach both
reflectivity edge and the main peak in the loss function in frames of EM theory
do not relate straightforwardly to the bare plasma frequency $\omega_p$.

\begin{figure}
\centering{\epsfig{file=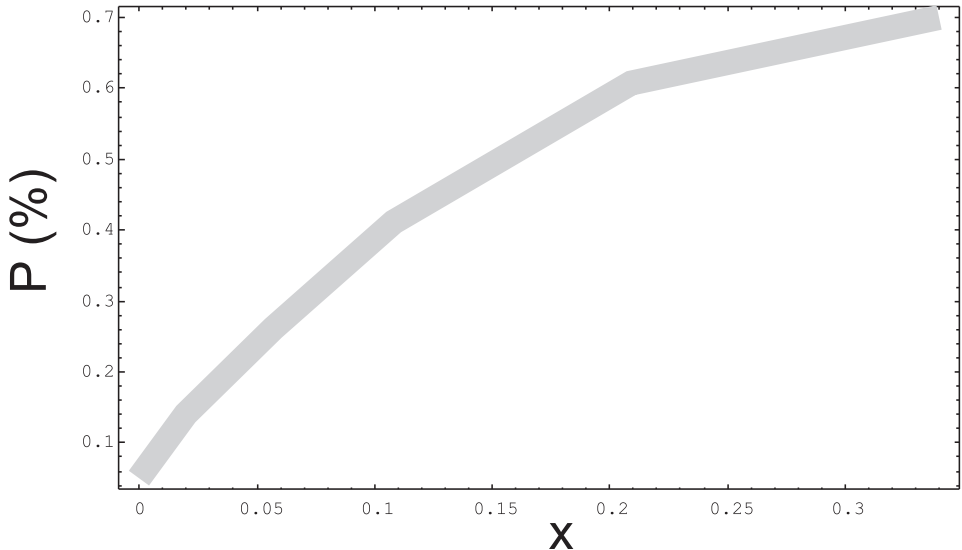,width=30pc}} \caption{Approximate dependence
of the metal volume fraction on Sr content in $La_{2-x}Sr_xCuO_4$.}
\label{PvsX}
\end{figure}

We assumed through the article that the doping level $x$ (say, strontium
content) can be simulated by an appropriate tuning of metal volume fraction $p$
in the composite. It is, however, desirable to establish the relation between
these quantities in more accurate manner. For the case of optical conductivity
in bulk single-crystalline  $La_{2-x}Sr_xCuO_4$ samples \cite{Uchida} this can
be done e.g. through the calculation relative decrease of the $2.2$ eV peak
with doping, then adjusting $p$ to keep the same proportion in theoretical
graph. The Fig. \ref{PvsX} illustrates  the relation $p\,(x)$ obtained in this
way. The saturation behaviour, discernible for large enough $x$ can be
attributed to the percolation phenomenon.

\section{Conclusion
}

In conclusion, we have proposed a new approach to theoretical description of
doped cuprates, treating these as inhomogeneous composite materials, containing
the dielectric and metallic stripe-like nanoparticles. The formalism of
effective medium theory is then applied for calculation of dielectric
permittivity and optical spectra of $La_{2-x}Sr_xCuO_4$ with $x$ varying in a
wide range. Reasonable quantitative agreement with experiment has been
obtained. The model in its simplest form was found able to reproduce all
essential features of the transmittance \cite{Suzuki}, optical conductivity
$\sigma(\omega)$, and EELS spectra \cite{Uchida}. Substantial difference in
spectral and doping dependence of optical conductivity for thin-film
\cite{Suzuki} and bulk samples \cite{Uchida} is easily explained if to only
assume different shape of metallic and dielectric regions in both materials.
New peaks in $\sigma(\omega)$ and absorption spectra, that emerge in the
midinfrared range upon doping are attributed to geometrical (Mie's) resonances.
Overall, the model theory allows to properly describe the evolution of optical
spectra at the percolative insulator-to-metal transition.

Finally, we would like to note that an occurrence of the phase separation with
percolative nature of the insulator-to-metal transition is accompanied by many
specific features which in many cases mask the real  electronic structure and
can lead to the erroneous conclusions. In particular, making use of either
experimental data as a trump in favor of either mechanism of the high-$T_c$
superconductivity should be made with some caution, if  a phase homogeneity of
the samples under examination is questionable. In this connection one may
mention a long-standing discussion regarding the nature of the midinfrared
bands in doped cuprates.

Further extension of the model developed here should include the size and shape
distribution of metallic-like inclusions, temperature effects, as well as
effects of different external factors such as pressure, isotopic substitution,
photo-doping. Of course, this semi-empirical model needs in more detailed
microscopic reasoning. \ack We are grateful to N. Loshkareva, Yu. Sukhorukov,
K. Kugel, P. Horsch for discussions. One of the authors (A.S.M.) acknowledges
stimulating discussion with N. DelFatti. The research described in this
publication was made possible in part by Award No.REC-005 of the U.S. Civilian
Research \& Development Foundation for the Independent States of the Former
Soviet Union (CRDF). The authors acknowledge a partial support from the Russian
Ministry of Education, grant E00-3.4-280.

\end{document}